\documentstyle[12pt]{article} 
\catcode`\@=11

\def\psfortextures{
\def\PSspeci@l##1##2{%
\special{illustration ##1\space scaled ##2}}}

\def\psfordvips{
\def\PSspeci@l##1##2{%
\d@my=0.1bp \d@mx=\drawingwd \divide\d@mx by\d@my%
\includegraphics{##1\space}}}

\def\psforoztex{
\def\PSspeci@l##1##2{%
\special{##1 \space
      ##2 1000 div dup scale
      \putsp@ce{\number-\psllx} \putsp@ce{\number-\pslly} translate}}}
\def\putsp@ce#1{#1 }

\def\psonlyboxes{
\def\PSspeci@l##1##2{%
\at{0cm}{0cm}{\boxit{\vbox to\drawinght
  {\vss
  \hbox to\drawingwd{\at{0cm}{0cm}{\hbox{(##1)}}\hss}
  }}}}}

\newdimen\drawinght\newdimen\drawingwd
\newdimen\psxoffset\newdimen\psyoffset
\newbox\drawingBox

\newread\epsffilein    
\newif\ifepsffileok    
\newif\ifepsfbbfound   
\newif\ifepsfverbose   
\newdimen\epsfxsize    
\newdimen\epsfysize    
\newdimen\epsftsize    
\newdimen\epsfrsize    
\newdimen\epsftmp      
\newdimen\pspoints     
\def\ReadPSize#1{
\edef\PSfilename{#1}
\global\def\epsfllx{72}
\global\def\epsflly{72}
\global\def\epsfurx{540}
\global\def\epsfury{720}
\openin\epsffilein=#1
\ifeof\epsffilein\errmessage{I couldn't open #1, will ignore it}\else
   {\epsffileoktrue \chardef\other=12
    \def\do##1{\catcode`##1=\other}\dospecials \catcode`\ =10
    \loop
       \read\epsffilein to \epsffileline
       \ifeof\epsffilein\epsffileokfalse\else
          \expandafter\epsfaux\epsffileline:. \\%
       \fi
   \ifepsffileok\repeat
   \ifepsfbbfound\else
    \ifepsfverbose\message{No bounding box comment in #1; 
using defaults}\fi\fi
   }\closein\epsffilein\fi
\def\psllx{\epsfllx}\def\pslly{\epsflly}%
\def\psurx{\epsfurx}\def\psury{\epsfury}%
\drawinght=\epsfury bp%
\advance\drawinght by-\epsflly bp%
\drawingwd=\epsfurx bp%
\advance\drawingwd by-\epsfllx bp%
}

{\catcode`\%=12 \global\let\epsfpercent=
\long\def\epsfaux#1#2:#3\\{\ifx#1\epsfpercent
   \def\testit{#2}\ifx\testit\epsfbblit
      \epsfgrab #3 @ @ @ \\%
      \epsffileokfalse
      \global\epsfbbfoundtrue
   \fi\else\ifx#1\par\else\epsffileokfalse\fi\fi}%

\def\epsfgrab #1 #2 #3 #4 #5\\{%
   \global\def\epsfllx{#1}\ifx\epsfllx\empty
      \epsfgrab #2 #3 #4 #5 @\\\else
   \global\def\epsflly{#2}%
   \global\def\epsfurx{#3}\global\def\epsfury{#4}\fi}%

\newcount\xscale \newcount\yscale \newdimen\pscm\pscm=1cm
\newdimen\d@mx \newdimen\d@my
\let\ps@nnotation=\relax
\def\psboxto(#1;#2)#3{\vbox{
   \catcode`\:=12
   \ReadPSize{#3}
   \divide\drawingwd by 1000
   \divide\drawinght by 1000
   \d@mx=#1
   \ifdim\d@mx=0pt\xscale=1000
         \else \xscale=\d@mx \divide \xscale by \drawingwd\fi
   \d@my=#2
   \ifdim\d@my=0pt\yscale=1000
         \else \yscale=\d@my \divide \yscale by \drawinght\fi
   \ifnum\yscale=1000
         \else\ifnum\xscale=1000\xscale=\yscale
                    \else\ifnum\yscale<\xscale\xscale=\yscale\fi
              \fi
   \fi
   \divide \psxoffset by 1000\multiply\psxoffset by \xscale
   \divide \psyoffset by 1000\multiply\psyoffset by \xscale
   \global\divide\pscm by 1000
   \global\multiply\pscm by\xscale
   \multiply\drawingwd by\xscale \multiply\drawinght by\xscale
   \ifdim\d@mx=0pt\d@mx=\drawingwd\fi
   \ifdim\d@my=0pt\d@my=\drawinght\fi
\message{[#3\space [BoundingBox\string:
\space\epsfllx\space\epsflly\space\epsfurx\space\epsfury]}%
\message{[scaled\space\the\xscale\string:
\space\the\drawingwd\space x \the\drawinght]]}%
 \hbox to\d@mx{\hss\vbox to\d@my{\vss
   \global\setbox\drawingBox=\hbox to 0pt{\kern\psxoffset\vbox to 0pt{
      \kern-\psyoffset
      \PSspeci@l{\PSfilename}{\the\xscale}
      \vss}\hss\ps@nnotation}
   \global\ht\drawingBox=\the\drawinght
   \global\wd\drawingBox=\the\drawingwd
   \baselineskip=0pt
   \copy\drawingBox
 \vss}\hss}
  \global\psxoffset=0pt
  \global\psyoffset=0pt
  \global\pscm=1cm
  \global\drawingwd=\drawingwd
  \global\drawinght=\drawinght
}}

\def\psboxscaled#1#2{\vbox{
  \catcode`\:=12
  \ReadPSize{#2}
  \xscale=#1
  \divide\drawingwd by 1000\multiply\drawingwd by\xscale
  \divide\drawinght by 1000\multiply\drawinght by\xscale
  \divide \psxoffset by 1000\multiply\psxoffset by \xscale
  \divide \psyoffset by 1000\multiply\psyoffset by \xscale
  \global\divide\pscm by 1000
  \global\multiply\pscm by\xscale
\message{[#2\space [BoundingBox\string:
\space\epsfllx\space\epsflly\space\epsfurx\space\epsfury]}%
\message{[scaled\space\the\xscale\string:
\space\the\drawingwd\space x \the\drawinght]]}%
  \global\setbox\drawingBox=\hbox to 0pt{\kern\psxoffset\vbox to 0pt{
     \kern-\psyoffset
     \PSspeci@l{\PSfilename}{\the\xscale}
     \vss}\hss\ps@nnotation}
  \global\ht\drawingBox=\the\drawinght
  \global\wd\drawingBox=\the\drawingwd
  \baselineskip=0pt
  \copy\drawingBox
  \global\psxoffset=0pt
  \global\psyoffset=0pt
  \global\pscm=1cm
  \global\drawingwd=\drawingwd
  \global\drawinght=\drawinght
}}

\def\psannotate#1#2{\def\ps@nnotation{#2\global\let\ps@nnotation=\relax}#1}
\def\pscaption#1#2{\vbox{
   \setbox\drawingBox=#1
   \copy\drawingBox
   \vskip\baselineskip
   \vbox{\hsize=\wd\drawingBox\setbox0=\hbox{#2}
     \ifdim\wd0>\hsize
       \noindent\unhbox0\tolerance=5000
    \else\centerline{\box0}
    \fi
}}}

\def\at#1#2#3{\setbox0=\hbox{#3}\ht0=0pt\dp0=0pt
  \rlap{\kern#1\vbox to0pt{\kern-#2\box0\vss}}}

\newdimen\gridht \newdimen\gridwd
\def\gridfill(#1;#2){
  \setbox0=\hbox to 1\pscm
  {\vrule height1\pscm width.4pt\leaders\hrule\hfill}
  \gridht=#1
  \divide\gridht by \ht0
  \multiply\gridht by \ht0
  \advance \gridht by \ht0
  \gridwd=#2
  \divide\gridwd by \wd0
  \multiply\gridwd by \wd0
  \advance \gridwd by \wd0
  \vbox to \gridht{\leaders\hbox to\gridwd{\leaders\box0\hfill}\vfill}}

\def\frameit#1#2#3{\hbox{\vrule width#1\vbox{
  \hrule height#1\vskip#2\hbox{\hskip#2\vbox{#3}\hskip#2}%
        \vskip#2\hrule height#1}\vrule width#1}}
\def\boxit#1{\frameit{0.4pt}{0pt}{#1}}

\catcode`\@=12 

\psfordvips

\jot = 1.5ex
\def\baselinestretch{1.65}
\parskip 5pt plus 1pt

\catcode`\@=11

\@addtoreset{equation}{section}

\def\@normalsize{\@setsize\normalsize{15pt}\xiipt\@xiipt
\abovedisplayskip 14pt plus3pt minus3pt%
\belowdisplayskip \abovedisplayskip
\abovedisplayshortskip  \z@ plus3pt%
\belowdisplayshortskip  7pt plus3.5pt minus0pt}
\def\small{\@setsize\small{13.6pt}\xipt\@xipt
\abovedisplayskip 13pt plus3pt minus3pt%
\belowdisplayskip \abovedisplayskip
\abovedisplayshortskip  \z@ plus3pt%
\belowdisplayshortskip  7pt plus3.5pt minus0pt
\def\@listi{\parsep 4.5pt plus 2pt minus 1pt
            \itemsep \parsep
            \topsep 9pt plus 3pt minus 3pt}}

\def\underline#1{\relax\ifmmode\@@underline#1\else
        $\@@underline{\hbox{#1}}$\relax\fi}
\@twosidetrue
\relax

\catcode`@=12

\evensidemargin 0.0in
\oddsidemargin 0.0in
\topmargin -0.2in
\textwidth 6.4in
\textheight 8.9in

\catcode`\@=11

\def\section{\@startsection{section}{1}{\z@}{3.5ex plus 1ex minus
   .2ex}{2.3ex plus .2ex}{\large\bf}}

\def\ps@headings{\def\@oddfoot{}\def\@evenfoot{}
\def\@oddhead{\hbox{}\hfill
        \makebox[.5\textwidth]{\raggedright\ignorespaces --\thepage{}--
        \hfill }}
\def\@evenhead{\@oddhead}
\def\subsectionmark##1{\markboth{##1}{}}
}

\ps@headings

\catcode`\@=12

\relax

\def\figcap{\section*{Figure Captions\markboth
        {FIGURECAPTIONS}{FIGURECAPTIONS}}\list
        {Fig. \arabic{enumi}:\hfill}{\settowidth\labelwidth{Fig. 999:}
        \leftmargin\labelwidth
        \advance\leftmargin\labelsep\usecounter{enumi}}}
 \relax
\def\tablecap{\section*{Table Captions\markboth
        {TABLECAPTIONS}{TABLECAPTIONS}}\list
        {Table \arabic{enumi}:\hfill}{\settowidth\labelwidth{Table 999:}
        \leftmargin\labelwidth
        \advance\leftmargin\labelsep\usecounter{enumi}}}
 \relax
\def\reflist{\section*{References\markboth
        {REFLIST}{REFLIST}}\list
        {[\arabic{enumi}]\hfill}{\settowidth\labelwidth{[999]}
        \leftmargin\labelwidth
        \advance\leftmargin\labelsep\usecounter{enumi}}}
 \relax

\catcode`\@=11

\def\marginnote#1{}

\def\ps@headings{\def\@oddfoot{}\def\@evenfoot{}
\def\@oddhead{\hbox{}\hfill
        \makebox[.5\textwidth]{\raggedright\ignorespaces --\thepage{}--
        \hfill }}
\def\@evenhead{\@oddhead}
\def\subsectionmark##1{\markboth{##1}{}}
}

\ps@headings

\relax

\def\firstpage#1#2#3#4#5#6{
\begin{document}
\begin{titlepage}
\nopagebreak
\title{\begin{flushright}
        \vspace*{-1.0in}
     {\normalsize NUB--#1 #2}\\[-9mm]
       {\normalsize hep-th/9811224}\\[14mm]
\end{flushright}
{#3}}
\author{\large #4 \\ #5}
\maketitle
\vskip -7mm     
\nopagebreak 
\def\baselinestretch{1.0}
\begin{abstract}
{\noindent #6}
\end{abstract}
\vfill
\begin{flushleft}
\rule{16.1cm}{0.2mm}\\[-3mm]
$^{\star}${\small Research supported in part by
the National Science Foundation under grant
PHY--96--02074.}\\
November 1998
\end{flushleft}
\thispagestyle{empty}
\end{titlepage}}
\newcommand{\Zint}{{\mbox{\sf Z\hspace{-3.2mm} Z}}}
\newcommand{\Real}{{\mbox{I\hspace{-2.2mm} R}}}
\def\simlt{\stackrel{<}{{}_\sim}}
\def\simgt{\stackrel{>}{{}_\sim}}
\newcommand{\dal}{\raisebox{0.085cm}
{\fbox{\rule{0cm}{0.07cm}\,}}}
\newcommand{\dt}{\partial_{\langle T\rangle}}
\newcommand{\dtbar}{\partial_{\langle\bar{T}\rangle}}
\newcommand{\al}{\alpha^{\prime}}
\newcommand{\mst}{M_{\scriptscriptstyle \!S}}
\newcommand{\mpl}{M_{\scriptscriptstyle \!P}}
\newcommand{\dv}{\int{\rm d}^4x\sqrt{g}}
\newcommand{\lv}{\left\langle}
\newcommand{\rv}{\right\rangle}
\newcommand{\ph}{\varphi}
\newcommand{\abar}{\bar{a}}
\newcommand{\sbar}{\,\bar{\! S}}
\newcommand{\xbar}{\,\bar{\! X}}
\newcommand{\fbar}{\,\bar{\! F}}
\newcommand{\zbar}{\bar{z}}
\newcommand{\dbar}{\,\bar{\!\partial}}
\newcommand{\tbar}{\bar{T}}
\newcommand{\taubar}{\bar{\tau}}
\newcommand{\ubar}{\bar{U}}
\newcommand{\tetabar}{\bar\Theta}
\newcommand{\etabar}{\bar\eta}
\newcommand{\qbar}{\bar q}
\newcommand{\ybar}{\bar{Y}}
\newcommand{\phb}{\bar{\varphi}}
\newcommand{\cm}{Commun.\ Math.\ Phys.~}
\newcommand{\prl}{Phys.\ Rev.\ Lett.~}
\newcommand{\pr}{Phys.\ Rev.\ D~}
\newcommand{\pl}{Phys.\ Lett.\ B~}
\newcommand{\ibar}{\bar{\imath}}
\newcommand{\jbar}{\bar{\jmath}}
\newcommand{\np}{Nucl.\ Phys.\ B~}
\newcommand{\F}{{\cal F}}
\renewcommand{\L}{{\cal L}}
\newcommand{\A}{{\cal A}}
\newcommand{\M}{{\cal M}}
\newcommand{\N}{{\cal N}}
\newcommand{\T}{{\cal T}}
\newcommand{\ads}{{\rm AdS}}
\renewcommand{\Im}{\mbox{Im}}
\newcommand{\e}{{\rm e}}
\newcommand{\be}{\begin{equation}}
\newcommand{\en}{\end{equation}}
\newcommand{\gsi}{\,\raisebox{-0.13cm}{$\stackrel{\textstyle
>}{\textstyle\sim}$}\,}
\newcommand{\lsi}{\,\raisebox{-0.13cm}{$\stackrel{\textstyle
<}{\textstyle\sim}$}\,}
\date{}
\firstpage{3192}{}
{\large\sc Remarks on Two-Loop Free Energy in $\N\,{=}\,\,4$ 
Supersymmetric\\[-5mm] 
 Yang-Mills Theory at Finite Temperature$^\star$}
{A. Fotopoulos and
T.R. Taylor}
{\normalsize\sl Department of Physics, Northeastern
University, Boston, MA 02115, U.S.A.}
{The strong coupling behavior of finite temperature free energy in
$\N\,{=}\,\,4$ supersymmetric $SU(N)$ Yang-Mills theory has been recently 
discussed  by Gubser, Klebanov and Tseytlin in the context of AdS-SYM
correspondence. In this note, we focus on the weak coupling behavior.
As a result of a two-loop computation we obtain, in the large $N$ 't Hooft limit,
$F(g^2N\to 0 )\approx -\frac{\pi^2}{6}N^2V_3T^4\,(1-\frac{3}{2\pi^2}g^2N)$.
Comparison with the strong coupling expansion 
provides further indication that free energy is a smooth monotonic 
function of the coupling constant.}
\setcounter{section}{0}

{}Finite temperature effects break supersymmetry \cite{gg}. By 
switching on non-zero
temperature one can interpolate between supersymmetric and non-supersymmetric
theories. For instance in gauge theories, one can interpolate between the 
supersymmetric case and a theory which contains pure Yang-Mills (YM) as the 
massless sector, with some additional thermal excitations.
In the infinite temperature limit, the time
dimension decouples and, at least formally, one obtains a 
non-supersymmetric Euclidean gauge theory.
If no phase transition occurs when the YM gas is heated up, then
the dynamics of realistic gauge theories such as QCD
is smoothly connected to their supersymmetric relatives.

Maldacena conjecture \cite{mald} which relates the large $N$ limit of 
$\N{=}\,4$ supersymmetric $SU(N)$ Yang Mills theory (SYM) to 
type IIB superstrings
propagating on $\ads_5\times S^5$, provides a very promising
starting point towards QCD. On the superstring side,
non-zero temperature can be simulated by including Schwarzschild
black holes embedded in AdS spacetime \cite{wit},
which describe the near-horizon geometry
of non-extremal D-brane solutions \cite{hs}. The classical geometry
of black holes with Hawking temperature $T$ does indeed encode correctly
many qualitative features of large $N$ gauge theory heated up
to the same temperature. At the quantative level though, the comparison
between SYM and supergravity becomes rather subtle because the supergravity
side merely provides  the strong coupling expansion for physical quantities
while most of finite temperature computations in SYM are limited
to the perturbative, weak coupling expansion. In this note, we comment
on the computation of free energy.

The SYM thermodynamics was first compared with the thermodynamics of
D-branes in ref.\cite{gkp}. The free energy $F$ obtained in ref.\cite{gkp}
describes the limit of infinitely strong coupled SYM theory.
More recently,
the AdS-SYM correspondence has been employed for computing
the subleading term in the strong coupling expansion 
(in $\lambda\equiv g^2N$)
\cite{gkt,ty}:\footnote{In this context, the gauge coupling $g$ is related to
the type IIB superstring coupling $g_s$: $g^2=2\pi g_s$.}
\begin{equation}
{}F(\lambda\to\infty) ~\approx~ -\frac{\pi^2}{6}N^2V_3T^4\,\big[\,\frac{3}{4}+
\frac{45}{32}\zeta(3)(2\lambda)^{-3/2}\,\big]\ .   \label{inf}
\end{equation}
The comparison with the limiting free-theory value,
\begin{equation}
{}F(\lambda=0 )~=~ -\frac{\pi^2}{6}N^2V_3T^4\ , \label{free}\end{equation}
indicates that the exact answer has the form:
\begin{equation}
{}F(\lambda)~=~-\frac{\pi^2}{6}N^2V_3T^4f(\lambda)\ ,\label{exact}\end{equation}
where the function $f(\lambda)$ interpolates smoothly between the asymptotic
values $f(0)=1$ and $f(\infty)=3/4$ \cite{gkp}. The sign of the subleading
correction ${\cal O}[(2\lambda)^{-3/2}]$ in eq.(\ref{inf})
indicates that $f$ decreases monotonically from 1 to 3/4.

The question whether free energy interpolates smoothly between
weak and strong coupling limits deserves careful investigation,
especially in view of the recent claim in favor
of a phase transition at finite $\lambda$ \cite{miao}.\footnote{
It is beyond the scope of this note to review the arguments of ref.\cite{miao},
however we would like to point out that they involve certain
assumptions on the convergence properties of perturbative expansions.
The proposed $2\pi^2$ convergence radius does not seem realistic
after one looks at the two-loop correction, see
eq.(\ref{two}).}
There is, however, a place to look for further hints on the
properties of free energy:
the subleading terms in the weak coupling expansion.
Surprisingly enough, they cannot be found in the existing literature. 
In order to fill this gap, we calculated the two-loop correction to
free energy. The result is:
\begin{equation}
{}F(\lambda\to 0 )~\approx~ -\frac{\pi^2}{6}N^2V_3T^4
\,[1-\frac{3}{2\pi^2}\lambda\, ]\ . \label{two}\end{equation}
The (relative) negative sign of the two-loop correction provides
further indication that the free energy is a smooth, monotonic
function of the 't Hooft coupling $\lambda$. In the following part 
of this note we present some details of the two-loop computation leading to
eq.(\ref{two}).

For the purpose of diagrammatic computations, it is convenient
to use the $\N\,{=}\,1$ decomposition of $\N\,{=}\,4$ SYM \cite{sym}, 
with one Majorana fermion corresponding to
the gaugino, and the three remaining Majorana fermions combined
with scalars in three $\N\,{=}\,1$ chiral multiplets.
The two-loop diagrams are displayed in Figure 1, together with
the combinatorial/statistics factors.
\begin{figure}
\[
\psannotate{\psboxto(0cm;5cm){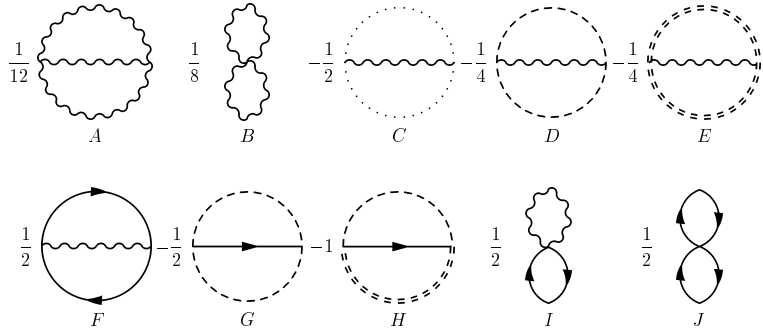}}{}
 \]\vskip -1cm
\caption{\em Two-loop diagrams contributing to free energy. Gauge
bosons are represented by wiggly lines, ghosts by dotted lines,
fermionic (Majorana) components of chiral multiplets by dashed lines,
scalars by solid lines and $\N\,{=}\, 1$ gauginos by double-dashed lines.}
\vskip 5mm
\end{figure}
The two-loop integrals can be readily performed by using
techniques described in refs.\cite{kap}. In the table below,
we list results for individual diagrams:\footnote{Diagrams
are computed in the Feynman gauge.}
\begin{equation}
\begin{array}{|c|c|c|c|c|c|c|c|c|c|}
\hline
{}~~A~~ & ~~B~~ & ~~C~~& ~~D~~& ~~E~~ &~~F~~ & ~~G~~ & ~~H~~& ~~I~~ 
&~~J~~\\ \hline
 -\frac{9}{4}\alpha& 3\alpha& \frac{1}{4}\alpha& \frac{3}{4}\beta& 
\frac{1}{4}\beta& 
-\frac{9}{2}\alpha&\frac{3}{2}\beta&\frac{3}{2}\beta& 12\alpha& 
\frac{15}{2}\alpha\\
\hline\end{array}\nonumber\end{equation}
where
\begin{eqnarray}
\alpha &=& g^2c_Ad\,V_3\bigg({T^4\over 144}+{T^2\over 12(2\pi)^3}\int
 {d^3\vec{k}\over  |\vec{k}|}\bigg)\ ,\\
\beta &=& g^2c_Ad\,V_3\bigg({5T^4\over 144}-{T^2\over 3(2\pi)^3}\int
 {d^3\vec{k}\over |\vec{k}|}\bigg)\ ,
\end{eqnarray}
with $d$ denoting the dimension of the gauge group and $c_A$ the Casimir
operator in the adjoint representation. Note that individual diagrams
contain ultraviolet divergences. After combining all
contributions, we obtain the (finite) result:
\begin{equation}
{}F_{\rm 2-loop}=16\alpha+4\beta=\frac{1}{4}g^2c_Ad\,V_3 T^4\ .\label{f2}
\end{equation}
Specified to the case of $SU(N)$, with $d=N^2-1$ and $c_A=N$,
in the leading large $N$ order the above result yields eq.(\ref{two}).

Finally, we would like to make a few remarks on the structure of higher-order
perturbative corrections. The computation of higher-order terms requires
reorganizing the perturbation theory to account for Debye screening
and yields terms non-analytic in
$\lambda$ such as ${\cal O}(\lambda^{3/2})$ and 
${\cal O}(\lambda^2\ln\lambda)$ \cite{kap,gpy}. The full ${\cal O}(\lambda^2)$
term requires a three-loop calculation \cite{arn} and a full accounting
of Debye screening at three loops would produce the ${\cal O}(\lambda^{5/2})$
terms. However, perturbation theory is believed to be incapable of
pushing the calculation to any higher order due to infrared problems 
associated with magnetic
confinement and the presence of non-perturbative ${\cal O}(\lambda^3)$
contributions \cite{kap,gpy}. 
It would be very interesting to analyze from this point of view
the strong coupling expansion.

\noindent {\bf Acknowledgments}

Most of this work was done while the authors were visiting
Laboratoire de Physique Th\'eorique et Hautes Energies
at l'Universit\'e Paris-Sud, Orsay.
We are grateful to Pierre Bin\'etruy and all members of LPTHE
for their kind hospitality.


\begin{thebibliography}{99}
\bibitem{gg} L. Girardello, M.T. Grisaru and P. Salomonson,
\np 178 (1981) 331.
\bibitem{mald} J. Maldacena, hep-th/9711200.
\bibitem{wit} E. Witten, hep-th/9803131.
\bibitem{hs} G.T. Horowitz and A. Strominger, \np 360 (1991) 197;\\
G.W. Gibbons and P.K. Townsend, \prl 71 (1993) 3754.
\bibitem{gkp} S.S. Gubser, I.R. Klebanov and A.W. Peet, \pr 54 (1996) 3915,
hep-th/9602135.
\bibitem{gkt} S.S. Gubser, I.R. Klebanov and A.A. Tseytlin, hep-th/9805156.
\bibitem{ty} A.A. Tseytlin and S. Yankielowicz, hep-th/9809032.
\bibitem{miao} Miao Li, hep-th/9807196.
\bibitem{sym} L. Brink, J. Schwarz and J. Scherk, \np 121 (1977) 77;\\
F. Gliozzi, J. Scherk and D. Olive, \np 122 (1977) 253.
\bibitem{kap} J.I. Kapusta, ``Finite Temperature Field Theory,''
Cambridge University Press (1989); \np 148 (1979) 461.
\bibitem{gpy} D. Gross, R. Pisarski  and L. Yaffe, Rev.\ Mod.\ Phys.\
53 (1983).
\bibitem{arn} P. Arnold and C. Zhai, \pr 50 (1994) 7603;
\pr 51 (1995) 1906.
\end{thebibliography}
\end{document}